\documentstyle[prl,twocolumn, aps]{revtex}
\begin{document}

\draft

\title{
Superconducting Gap, Pseudogap and Their Relationship Investigated by Break-junction Tunneling Spectroscopy on $Bi_2Sr_2CaCu_2O_{8+\delta}$ Single Crystals
}

\author{
Y. Xuan$^1$, H. J. Tao$^1$\cite{responce}, Z. Z. Li$^1$, C. T. Lin$^2$, Y. M. Ni$^1$, B. R. Zhao$^1$, Z. X. Zhao$^1$
}
\address{
$^1$National Laboratory for Superconductivity,
Institute of Physics and Center for Condensed Matter Physics,
Chinese Academy of Sciences, P.O. Box 603, Beijing 100080, China\\
$^2$Max-Planck-Institut f\"{u}r Festk\"{o}rperforschung, 70569 Stuttgart, Germany
}

\maketitle

\begin{abstract}
Temperature and doping dependent tunneling spectroscopy on $Bi_2Sr_2CaCu_2O_{8+\delta}$ single crystals has been investigated by using break-junction technique. The results provide direct evidence that the pseudogap persists into heavily overdoped regime, possibly in whole doping range of the superconducting phase. The gap value determined from the peak-to-peak of the conductance, $\Delta_{p-p}$, exhibits a jump-like increase as $T$ is raised across the critical temperature. The magnitude of the jump, the pseudogap opening temperature and superconducting gap decrease with raising doping. The $\Delta_{p-p}$ below $T_c$ can be naturally treated as the vector sum of two competing orders: a BCS-like superconducting gap and a residual pseudogap.
   
\end{abstract}

\pacs{PACS numbers: 74.50.+r, 74.25.-q, 74.25.Dw, 74.72.Hs}

It is generally agreed that for underdoped cuprate high-$T_c$ superconductors  a generic phenomenon is there is the pseudogap (PG) in their excitation spectrum for temperature ($T$) above the critical temperature $T_c$. The PG has been observed with various probes both spin and charge sensitive\cite{Timusk}. Among the probes tunneling spectroscopy, one of the direct measures of quasiparticle excitations in a superconductor has attracted great deal of attention. However, thus far the results of tunneling, are still controversial. Different types of relation between the SG and the PG have been reported, such as smooth evolution\cite{Renner}, competing\cite{Ekino}, coexistence of the two gaps\cite{Krasnov}, and distinct two gaps\cite{Suzuki}, with scanning tunneling microscope (STM)\cite{Renner}, break-junction\cite{Ekino}, and c-axis intrinsic junction\cite{Krasnov,Suzuki}. There is also the question that whether the PG exists in overdoped regime, especially in heavily overdoped region? Apparently, experimentally clarifying the above discrepancies is a pressing task and will set constraint to any theoretical models on the origin of the pseudogap and on the mechanism of high-$T_c$ superconductivity. 

    In this paper we present systematic investigation on the detailed evolution behavior of the gap $\Delta_{p-p}$, determined by a quarter of the span from the peak-to-peak in the tunneling conductance, on $Bi_2Sr_2CaCu_2O_{8+\delta}$ (Bi2212) single crystals with break-junction technique, specifically focusing on the overdoped regime. For all samples studied, including the highly overdoped ($T_c$ = 72 K), our low-noise tunneling conductance clearly demonstrates the existence of the pseudogap. Detailed $T$-dependent measurements show a jump-like increase of $\Delta_{p-p}$ as $T$ is raised across $T_c$ within a very narrow temperature range, indicating an unsmooth transformation from the superconducting state to the PG state. We show that the measured gap for $T$ below $T_c$ can be naturally treated as a vector sum of two competing components: the superconducting gap (SG) and the residual PG. The PG opening temperature, $T^*$, the magnitude of the jump of $\Delta_{p-p}$ at $T_c$, $\Lambda$, and the value of the SG, and PG decrease with increase of doping. The doping dependence of $T^*$ implies that for Bi2212 the PG may exist in the whole overdoped regime provided there is the superconductivity. To our knowledge, this doping dependence of $\Lambda$, as a characteristic parameter for the unsmooth transformation is newly observed by tunneling, and the present work is the first directly tunneling observation of the doping dependence of the $T^*$.

    We chose break-junction to carry out our tunneling investigation because it provides a clean and fresh interface easily, which avoids surface degradation and diminishes the barrier scattering. In addition, the break-junction is of the superconductor-insulator-superconductor (SIS) type junction, which enhances the gap feature in the tunneling conductance due to its unique convolution relationship from the electronic density of states (DOS) of two superconductors on both sides of tunnel barrier. Therefore, the peak position is less affected by thermal smearing and can be taken as a characteristic parameter\cite{Wolf}. More importantly, previous studies at 4.2 K have provided reproducible and low-noise tunneling conductance consisting of sharp coherence peaks, a close to zero zero-bias conductance, and a significant peak-dip-hump structure\cite{Mandrus,Miyakawa,Zasadzinski}. Those facts demonstrate that the break-junction technique is a reliable tool with high-energy resolution to investigate the tunneling spectroscopy. 

    The Bi2212 single crystals used in this study were of three doping levels. The near-optimally doped samples were the as-grown crystals grown by traveling-solvent-floating-zone method. The medium-overdoped samples were achieved by annealing the as-grown crystals ($T_c$ = 89 K), synthesized by directional solidification method, in flowing O$_2$ at 500$^\circ$C for 120 hours. And the heavily overdoped samples, also from the as-grown Bi2212 single crystals grown by directional solidification, were obtained by annealing them under a pressure of 140 atm of O$_2$ at 500$^\circ$C for 50 hours. The on-set $T_c$ of the three batches of samples was 90 K, 82 K and 72 K with transition width of 1 K, 1 K and 3 K, respectively, measured by AC susceptibility using Lakeshore AC Susceptometer-7000. Hereafter we label the samples as OP90, OV82 and OV72. All the characterizations were carried out on segments of the very single crystals used to measure tunneling spectroscopy. The tunneling conductance curve, $dI/dV - V$, was measured by using a standard low-frequency ac lock-in technique with the four-probe method and a computer data acquisition system. To achieve low-resistive contacts ($<$ 2 $\Omega$) between the crystals and the leads, gold film electrodes were sputtered onto each sample.

    The break-junctions were formed in situ at 4.2 K\cite{Mandrus}. To realize it, thin (less than 0.1 mm) single crystal was glued onto phosphor-copper substrate with epoxy, and broken by applying a bending force perpendicular to the CuO$_2$ plane with a precise differential-screw. After a break-junction was formed, a key problem left was to keep it from being unstable as temperature varied. Because there is mismatch between the thermal expansion coefficients among the Bi2212 crystal, epoxy and phosphor-copper substrate, which often causes the contact of two pieces of the broken off crystal become loosen or tighten, and leads to the junction conductance changed drastically when the temperature varies. It even breaks down the junction at high temperature. Through practice we found that by adjusting the thickness of epoxy, which depends on the dimensions of the substrate, we were able to overcome this problem and perform stable $T$-dependent investigation. 

    To diminish the complexity bringing about by Josephson current and to facilitate the analysis of experimental data, each time we carefully adjusted the differential screw to increase the junction resistance and depress the Cooper pair tunneling to undetectable small. Tunneling measurements were carried out at temperatures ranging from near 4.2 K to where the PG disappears.

     Figure 1, 2 and 3 show the normalized tunneling conductance of different temperatures measured on OP90, OV82 and OV72, respectively. The normalization was done by divided each measured conductance by its conductance of high bias voltage. For clarity, the curves of $T \leq T_c$ are sequentially shifted up with a unit value (in (a) panel of each figure), whilst those of $T \geq T_c$ are plotted with same coordinate scale as they were measured (in (b) panel of each figure). 

    In Fig. 1 to 3 several points are notable. (1) At low $T$, the conductance for OP90 and OV82 exhibit the significant dip-hump structure, which has been observed in ARPES\cite{Ding} and tunneling measurements and is related to the ($\pi$, $\pi$) resonance in inelastic neutron scattering\cite{Zasadzinski}. The dip feature for OV72 is not as significant as of OP90 and OV82, but it is still visible in Fig. (3.c) by comparing the low-$T$ conductance to the ones of higher temperature. Those facts demonstrate the high quality of our junctions. (2) For $T \ll T_c$, such as $T$ = 4.2 K, it is generally safe to take $\Delta_{p-p}$ as the SG, $\Delta$(0). The low-T parameters, including $\Delta$(0), and the dip position $E_{dip}$ in tunneling conductance for the three samples are listed in Table I. These values are consistent well with the results of Ref. \cite{Zasadzinski}, in which a detailed examination over a wide range of oxygen doping at 4.2 K was performed. On the other hand, the tunneling curve calculated from a weak-coupling d-wave DOS has significant deviation from the experimental date in the shape of conductance peak\cite{Zasadzinski,Coffey}. This brings about difficulty to model fit the experimental curves even in the low $T$. (3) In Fig. 1, 2 and 3, the conductance curves at $T > T_c$ clearly demonstrate the existence of PG even in heavily overdoping level with the presence of peaks, and the low-noise background enables us to get the value of $\Delta_{p-p}(T)$, a quarter of the peak-to-peak span. (4) The opening temperature of the PG, $T^*$, can be determined as the temperature at which the conductance peaks vanish, meanwhile the central dip in the conductance is filled-in. By so doing, the obtained $T^*$ is 190, 150 and 120 K for sample OP90, OV82 and OV72, respectively, and we estimate the error bars are $\pm$15, $\pm$15, $\pm$10 K, respectively. 

    To gain more insights into the physics, we depict in Fig. 4 the $\Delta_{p-p}(T)$ taken from Fig. 1, 2 and 3, respectively, as a function of the normalized temperature $T/T_c$,. The filled squares, deltas and circles is corresponding to sample OP90, OV82 and OV72, respectively. The solid lines are the BCS $T$-dependent gaps corresponding to measured $\Delta$(0) of OP90, OV82 and OV72, respectively. A general feature for the three junctions in Fig. 4 is that for $2/3T_c < T < T_c$, the $\Delta_{p-p}(T)$ decreases with increasing $T$ much slower than the BCS behavior. There are at least two factors that may cause this slow decrease. First, the life-time effect of the quasuiparticles, which smears the tunneling conductance and gives rise to a larger $\Delta_{p-p}(T)$ than the SG, $\Delta_s(T)$. Second, the measured single gap is a vector sum of the SG and the residual PG for $T$ below $T_c$: $\Delta_{p-p}(T) = (\Delta_s^2(T) + \Delta_p^2(T))^{1/2}$. For simplicity, if we do not take into account the life-time broadening effect, and subtract the BCS gap from the $\Delta_{p-p}(T)$ according to the above formula we get the residual PG for $T < T_c$, which is shown in the inset of Fig. 4. The data for $T \geq T_c$ in the inset are the same as those in the main part of Fig. 4. The vector sum scenario is suggested by a number of theoretical models\cite{Chakravarty} stating that in superconducting state there exist two competing orders: the SG and the PG. As temperature is raised across $T_c$, $\Delta_{p-p}$ undergoes an abrupt jump within a very narrow temperature range ($\sim$2 K in OP90) and the $T_c$ can be regarded as where the maximum of $d\Delta_{p-p}(T)/dT$ is, or the middle of the jump. This behavior demonstrates that the maximum of DOS moves to higher energy unsmoothly as soon as long-range phase coherence disappears. To characterize this unsmooth behavior, we introduce $\Lambda$, the jump in $\Delta_{p-p}$ within the measured temperature interval in the vicinity of $T_c$. It can be seen that $\Lambda$ has the same trend with doping as $T^*$ (see also Tab. I).

    Here one point we would like to emphasis is that the measured gap $\Delta_{p-p}$ in the {\it normal} state is not the true magnitude of the PG, because apparently the value of the PG should decrease or keep as constant when $T$ is raised. In general, high fluctuation\cite{Dynes} or strong pair-breaking\cite{Burstein} can bring out broadening conductance peak and widen separation between the peaks in the conductance. As to how to extract the true value of the PG and its temperature behavior from tunneling date is an open issue remaining to be address. The fact that for OP90 $\Delta_{p-p}$ keeps increasing as $T$ is raised reflects that in the normal state there is high fluctuation or strong pair-breaking which enhance as $T$ is raised. Moreover, the trend of doping dependence of $\Delta_{p-p}(T)$ demonstrates that the fluctuation or pair-breaking effect become weak in high doping region. Those features are consistent with the picture of precursor paring\cite{Emery} in electronic or spin channel\cite{Anderson} in the normal state but we emphasize it exists in the overdoped regime, and directly show that the difference in electronic excitation between the superconducting state and the normal state near $T_c$ increases with lowering doping.     

To further illustrate the present observation we plot the doping dependence of the $\Delta(0)$ and $T^*$ in Fig. 5, in which the dotted line is the BCS mean-field gap, $\Delta = 2.14k_BT_c$, the hole concentration $p$ is obtained from the empirical relation $T_c/T_{c,max} = 1- 82.6 (p - 0.16)^2$\cite{Tallon1}, with $T_{c,max} = 92 K$. The filled circles are measured $\Delta$(0) and the filled squares are the measured $2.14k_BT^*$. The thick straight line is a guide to the eye. It is surprising that the extrapolation of the thick line just crosses to the abscissa at $p$ = 0.27, where the superconductivity disappears. That implies so long as it is a superconductor, there may be the PG state. A similar trend for Bi2212 can be found in Ref.\cite{Oda}. This phase diagram of Bi2212 is qualitatively different from that of YBa$_2$Cu$_3$O$_{7-\delta}$, to which it is shown the PG state only exists in the region where the hole concentration $p$ is less than a certain value, 0.19\cite{Tallon2}. Obviously this is an issue that needs to be further addressed. 

    To conclude, we have performed investigations on temperature and doping dependent tunneling spectroscopy of Bi2212 single crystals with break-junction technique, focusing on overdoped regime. The results provide direct evidence that for Bi2212 the PG persists into heavily overdoped regime, possibly in whole doping range accompanying the SG. Detailed measurements near $T_c$ have showed the measured $\Delta_{p-p}$ undergoes a jump-like increase within a very narrow temperature range as $T$ is raised across $T_c$, indicating an unsmooth transformation from the superconducting state to the PG state. The magnitude of the jump of $\Delta_{p-p}$, $T^*$, the SG, and PG decrease with raising doping. We show the measure gap $\Delta_{p-p}(T)$ for $T$ below $T_c$ can be naturally treated as the vector sum of two competing orders: a BCS-like SG and a residual PG. 

\acknowledgements
This work was supported by a grant for State Key Program for Basic Research of China, and the National Center for Research and Development on Superconductivity.

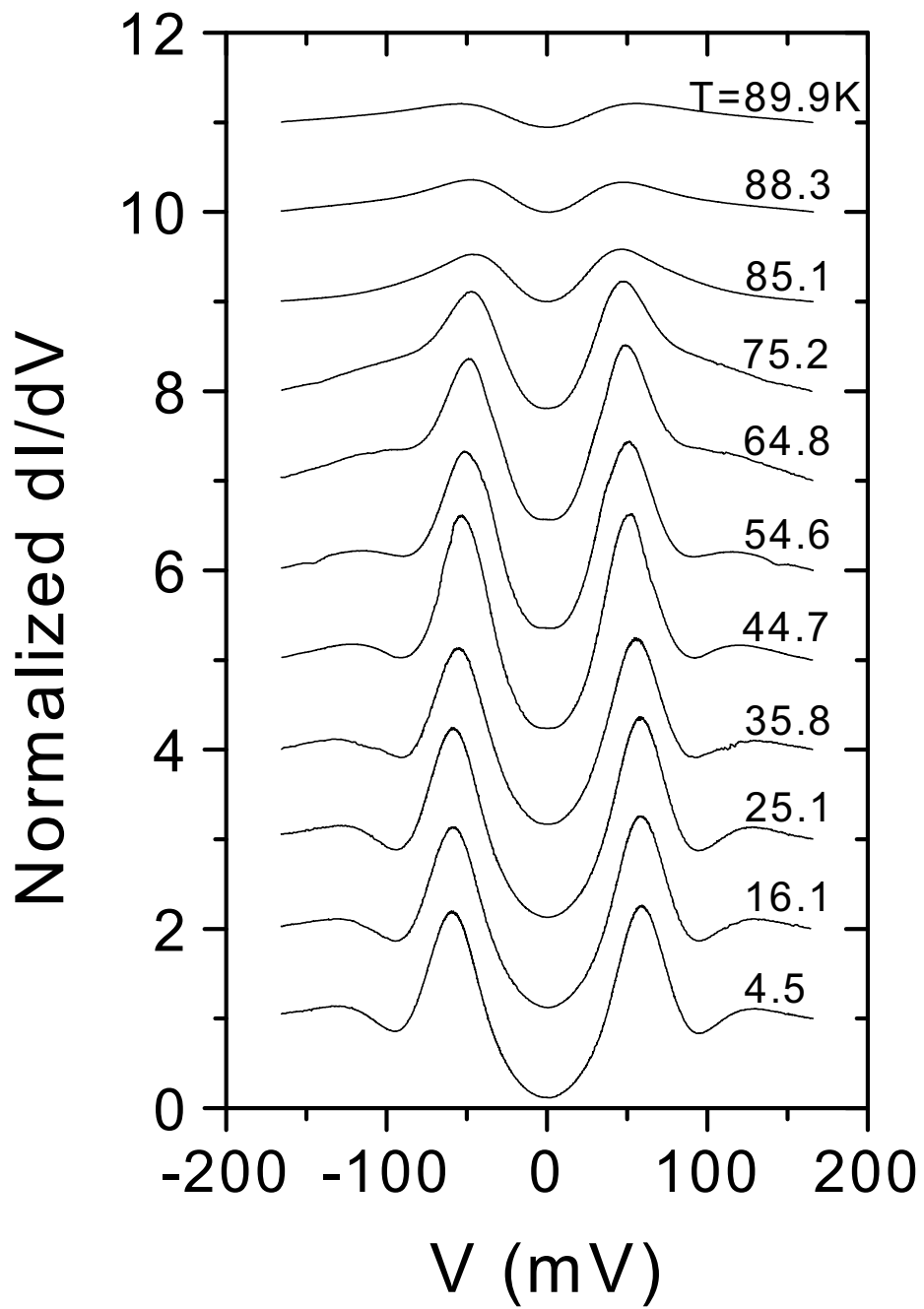
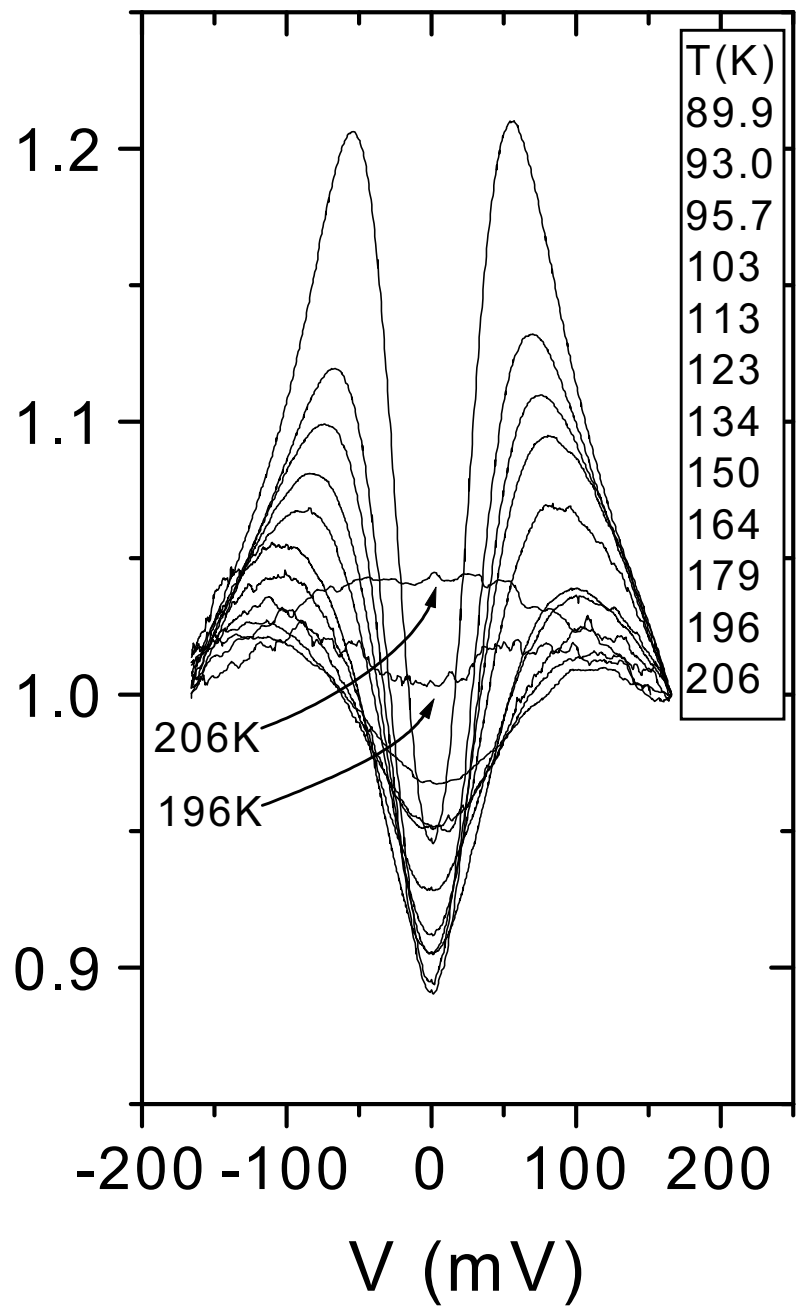
\begin{figure}
\caption{
$T$-dependence of SIS tunneling conductance on OP90 (near-optimally doped Bi2212) break-junction. Each curve has been normalized by condutance value of highest positive biased voltage. For clarity, (a) curves for $T \leq T_c$ are sequentially shifted up with a unit value; (b) curves for $T \geq T_c$ are plotted with same coordinate scale as they were measured.
}
\end{figure}
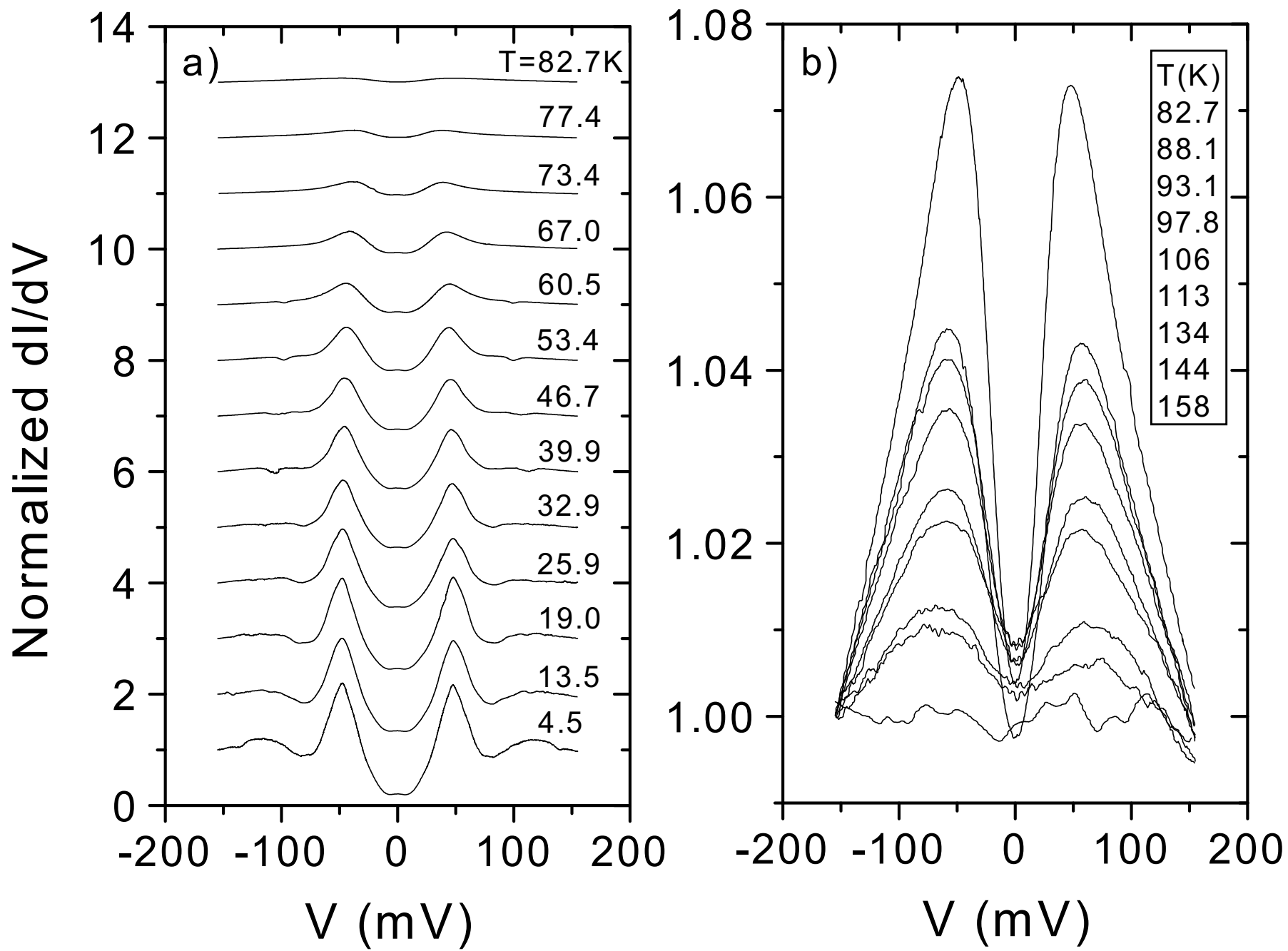
\begin{figure}
\caption{
$T$-dependence of SIS tunneling conductance on OV82 (medium overdoped Bi2212) break-junction. Each curve has been normalized by conductance value of highest negative biased voltage. For clarity, (a) curves for $T \leq T_c$ are sequentially shifted up with a unit value; (b) curves for $T \geq T_c$ are plotted with same coordinate scale as they were measured.
}
\end{figure}
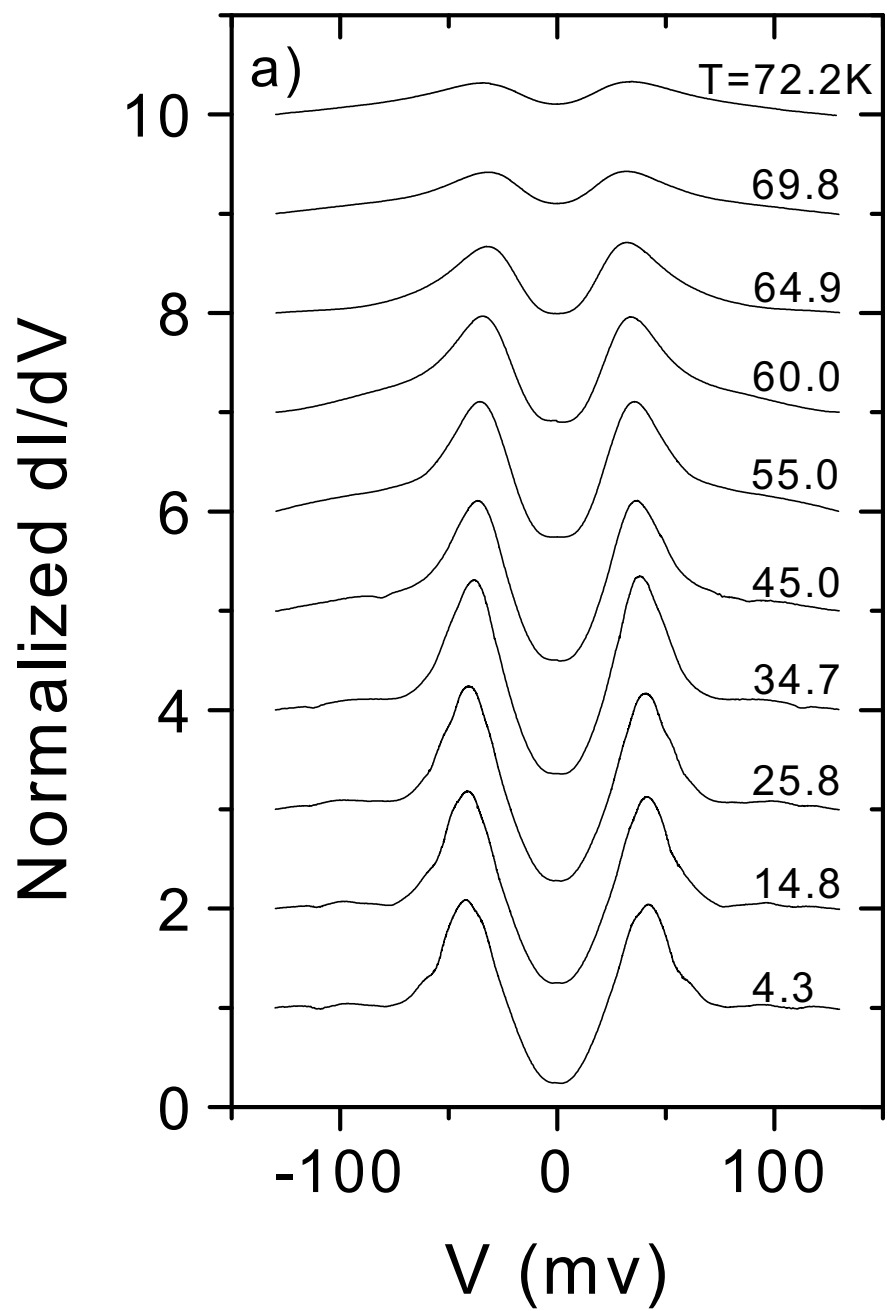
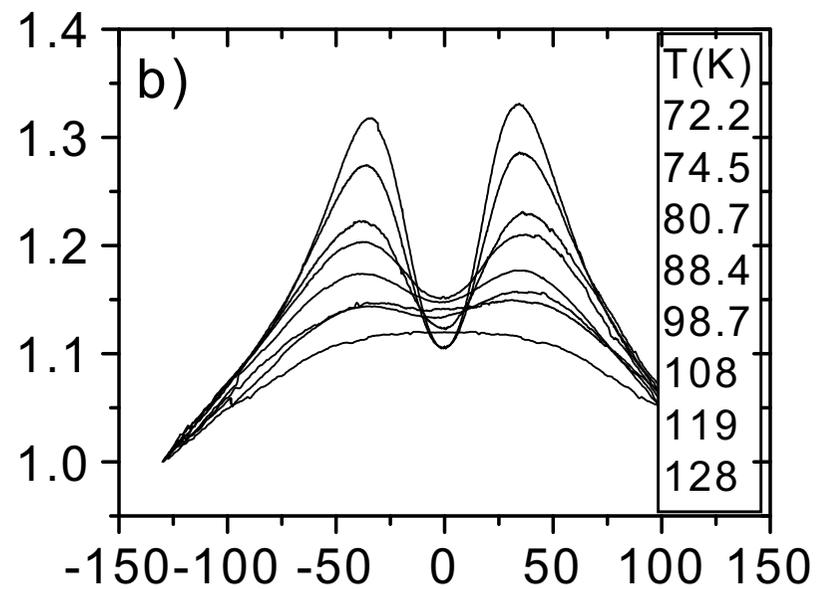
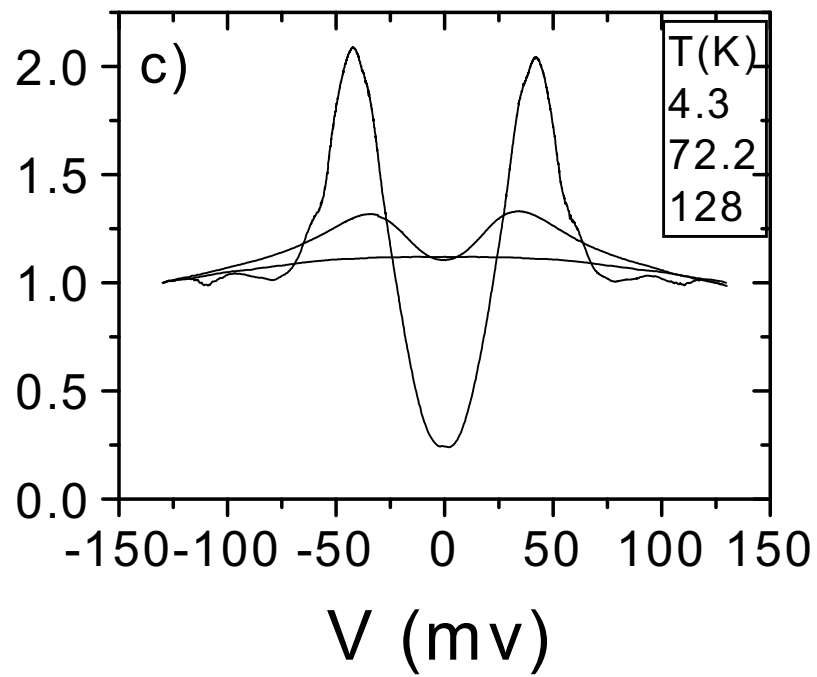
\begin{figure}
\caption{
$T$-dependence of SIS tunneling conductance o on OV72 (heavily overdoped Bi2212) break-junction. Each curve has been normalized by conductance value of highest negative biased voltage. For clarity, (a) curves for $T \leq T_c$ are sequentially shifted up with a unit value; (b) curves for $T \geq T_c$ are plotted with same coordinate scale as they were measured.
}
\end{figure}
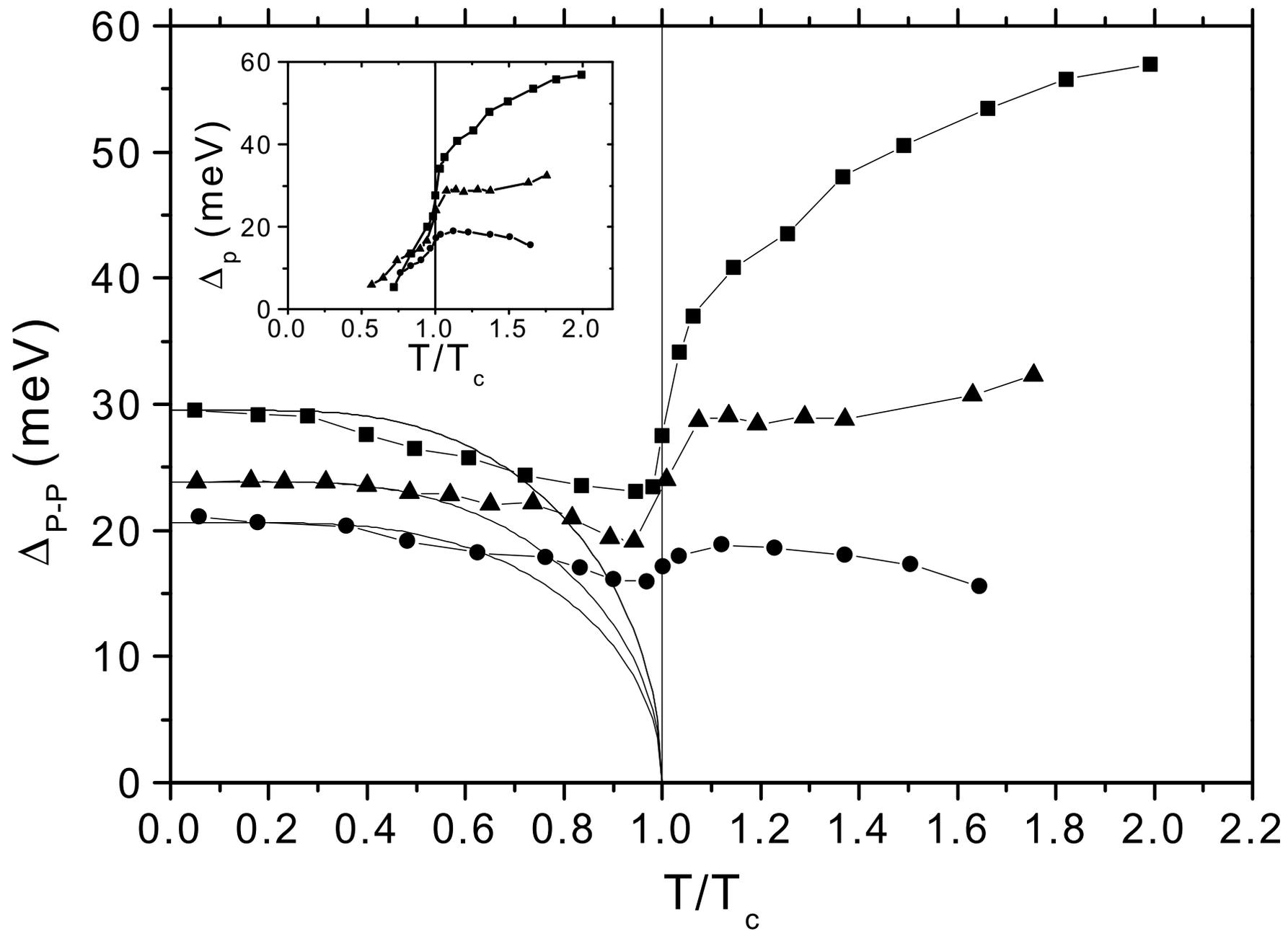
\begin{figure}
\caption{
$T$-dependence of $\Delta_{p-p}s$ of OP90 (square), OV82 (triangle), OV72 (circle) from Fig. 1, 2 and 3, respectively, with the normalized temperature, $T/T_c$. For comparison, BCS $T$-dependence SG, $\Delta_s(T)$, is plotted for OP90, OV82, OV72 in solid line, respectively. Inset shows the $\Delta_p$ from the separation as $\Delta_{p-p} = (\Delta_s^2 + \Delta_p^2)^{1/2}$, with the normalized temperature, $T/T_c$.}
\end{figure}
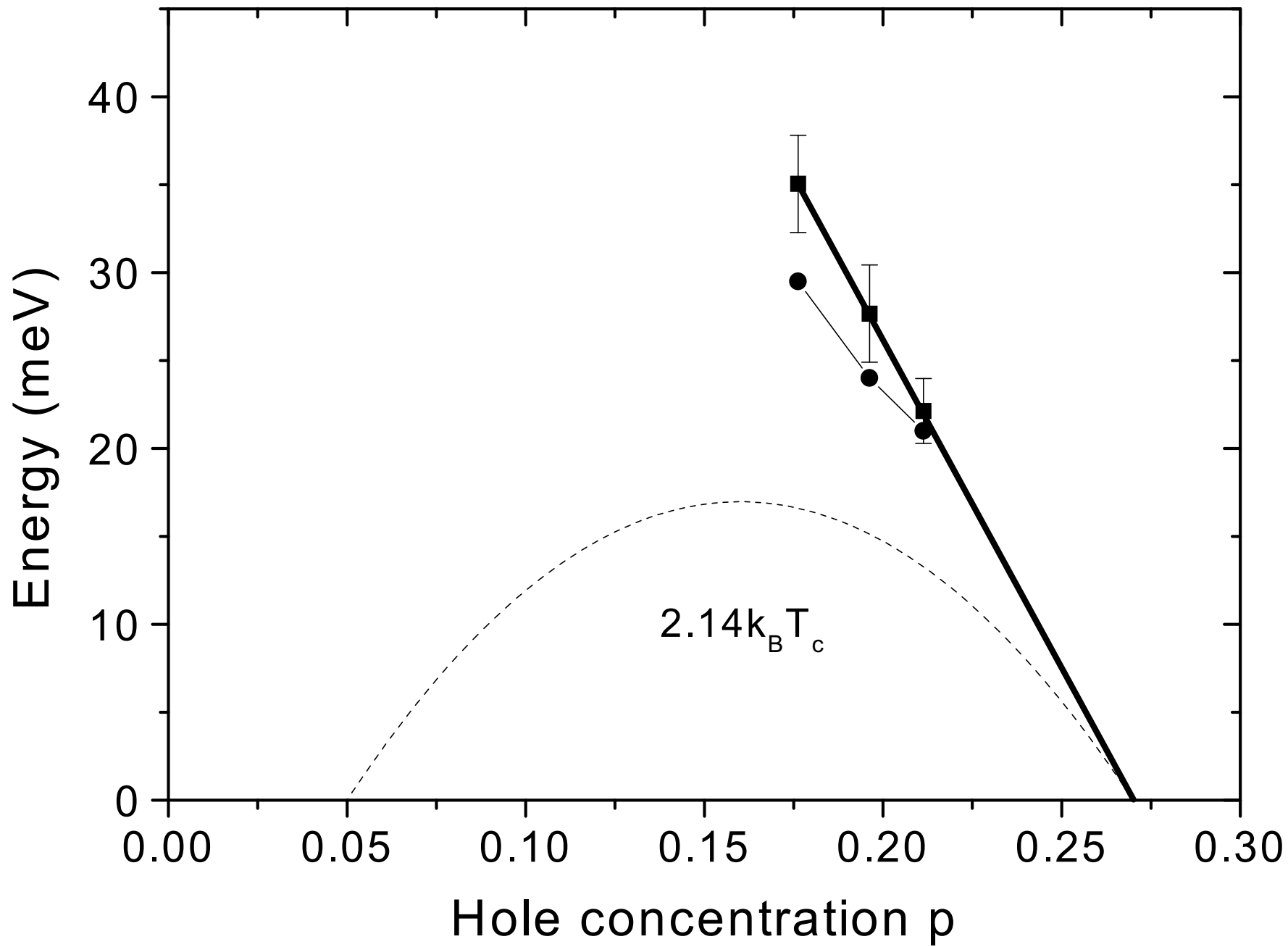
\begin{figure}
\caption{
Plot of $\Delta$(0) (circle) and $2.14k_BT^*$ (square) vs hole concentration $p$. The thick straight line is a guide to the eye, and its extrapolation crosses to the abscissa just at the superconducting phase boundary. The dashed line is the plot of BCS d-wave gap which is equal to $2.14k_BT_c$.
}
\end{figure}


\begin{references}
\bibitem[*]{responce} Email address: hjtao@aphy.iphy.ac.cn
\bibitem{Timusk} T. Timusk and B. Statt, Rep. Prog. Phys. {\bf 62}, 61 (1999).
\bibitem{Renner} C. Renner et al., Phys. Rev. Lett. {\bf 80}, 149 (1998). 
\bibitem{Ekino} T. Ekino, Y. Sezaki, and H. Fujii, Phys. Rev. B {\bf 60}, 6916 (1999). 
\bibitem{Krasnov} V. M. Krasnov et al., Phys. Rev. Lett. {\bf 84}, 5860 (2000); V. M. Krasnov, A. E. Kovalev, A. Yurgens, and D. Winkler, Phys. Rev. Lett. {\bf 86}, 2657 (2001). 
\bibitem{Suzuki} M. Suzuki and T. Watanabe, Phys. Rev. Lett. {\bf 85}, 4787 (2000). 
\bibitem{Wolf} E. L. Wolf, {\it Principles of Electron Tunneling Spectroscopy}, (Oxford University, New York, 1985).
\bibitem{Mandrus} D. Mandrus, L. Forro, D. Koller, and L. Mihaly, Nature {\bf 351}, 460 (1991)
\bibitem{Miyakawa} N. Miyakawa et al., Phys. Rev. Lett. {\bf 80}, 157 (1998); N. Miyakawa et al., Phys. Rev. Lett. {\bf 83}, 1018 (1999).
\bibitem{Zasadzinski} J. F. Zasadzinski et al., Cond-mat/0102475. 
\bibitem{Ding} H. Ding et al., Phys. Rev. Lett. {bf 76}, 1533 (1996).
\bibitem{Coffey} L.Coffey, Cond-mat/0103518.
\bibitem{Chakravarty} S. Chakravarty, R. B. Laughlin, D. K. Morr, and C. Nayak, Phys. Rev. B {\bf 63}, 094503 (2001); I. Kosztin, Q. J. Chen, Y. J. Kao, and K. Levin, Phys. Rev. B {\bf 61}, 11662 (2000); J. W. Loram et al., J. Supercond. {\bf 7}, 243 (1994); R. S. Markiewicz, C. Kusko, and V. Kidambi, Phys. Rev. B {\bf 60}, 627 (1999). 
\bibitem{Dynes} R. C. Dynes, V. Narayanamurti, and J. P. Garno, Phy. Rev. Lett. {\bf 41}, 1509 (1978); A. A. Varlamov, G. Balestrino, E. Milani, and D. V. Livanov, Adv. Phys. {\bf 48}, 655, (1999).
\bibitem{Burstein} E. Burstein and S. Lundqvist, {\it Tunneling Phenomena in Solids}, (Plenum, New York, 1969), Chapter 29, P. Fulde; Chapter 30, T. Claeson.
\bibitem{Emery} V. J. Emery and S. A. Kivelson, Nature {\bf 374}, 434 (1995).
\bibitem{Anderson} P. W. Anderson, Science {\bf 235}, 1196 (1987).
\bibitem{Tallon1} J. L. Tallon et al., Phys. Rev. B {\bf 51}, 12911 (1995).
\bibitem{Oda} M. Oda, T. Matsuzaki, N. Momono, and M. Ido, Physica C {\bf 341-348}, 847 (2000).
\bibitem{Tallon2} J. T. Tallon and J. W. Loram, Physica C {\bf 349}, 53 (2001).

\end{references}
\end{document}